# A Two-Stage Stochastic Programming Model for Blood Supply Chain Management, Considering Facility Disruption and Service Level

Mohammad Arani[a,*], Mohsen Momenitabar[b], Zhila Dehdari Ebrahimi[b], Xian Liu[a]

[a]University of Arkansas at Little Rock, Department of Systems Engineering, Little Rock, USA
[b]North Dakota State University, Department of Transportation, Logistics and Finance, Fargo, USA

* Corresponding author: *E-mail address*: mxarani@ualr.edu.

**Abstract**

In this paper, a blood supply chain network, where the occurrence of disruption might interrupt the flow of Red Blood Cells, is dealt with. In principle, the probability of disruption is not the only property confiding the network, but unprecedented fluctuations in supplies and demands also contribute to the network's shortages and outdated blood units. Although the consideration of parameter uncertainties is of paramount importance in the real-world circumstances for a decision-maker, she or he would be willing to monitor the network in a properly broader perspective. Therefore, one of the eminent key performance indicators known as service level turned our attention. To tackle uncertainties in the mentioned network - comprising of the four conventional levels containing donors, blood collection facilities, blood banks, and hospitals - we present a two-stage stochastic programming model. Consequently, a toy-example is randomly generated to validate the proposed model. Furthermore, numerical analysis led us to a comprehensive service level analysis. Finally, potential pathways for future research are suggested.

## 1. Introduction

Having a reliable network of blood supplies is particularly important in case of medical emergencies. Blood Supply Chain Management (BSCM) generally comprises blood components collection, screening, processes, preserving, distribution, and medical procedures. Red Blood Cells (RBCs), platelets, Cryoprecipitate, and plasma are the most remarkable components of blood in terms of planning for BSCM. These blood components differ in shelf lives and storage conditions, where platelets can only be used for five days after collection, and forty-two days are considered for preserving RBCs, for instance. The perishability of blood components is on one hand and the interchangeability of products, which is called crossmatching, on the other hand, makes it extremely challenging to regulate the BSC.

One pillar of a sophisticated healthcare system is built upon an effective BSCM. Blood supplies are, indeed, the most important part of the medical infrastructure, which helps to save people's lives from chronic diseases such as cancer, severe trauma, and leukemia on a daily basis. BSCM aims at providing safe and sufficient blood supplies despite the high degree of complexity. Although the associated cost of blood supplies in the level of donors could be inconsiderable, since the voluntary blood donations are the significant sources of blood, screening, processing, preserving, and medical procedures pertinent to the rest of the BSCM are daunting expensive tasks that people lives strongly depend on them.

One of the primary blood components is the RBCs unit. RBCs units solely account for the flow of the product of more than sixty-four percentage over the distribution network [1]. Mainly, RBCs are divided into eight categories: A, B, AB, and O with the absence or presence of a protein called Rhesus (Rh). One medical property of RBCs is the crossmatching, in other words, it is of paramount importance to administer substitution type by observing alternatives as far as advised medically. While substitution brings the opportunity of a fewer number of shortages, the network would encounter a fewer number of outdated units as well [2,3], Table 1 presents the crossmatching rules.

Although the consideration of blood types and crossmatching add to the complexity of BSCM, unintended disruptions like the natural hazards in blood collection facilities and blood banks where they are responsible for preserving and advanced screening are of deep interest to managers. In case of a disruption, the BSC encounters adverse effects such as inventory shortages and delay in responding orders. In this paper, the uncertainties in supplies and demands are considered along with the possibilities of disruption in blood collection and blood bank facilities. To address the concern of a decision-maker, we selected a key performance indicator (KPI) called service level to properly observe the BSC. Moreover, we formulate the problem in the form of two-stage stochastic programming with two objectives of minimizing the total cost of network and maximizing the service level.

Table 1. RBCs cross-matching matrix

| Recipient | | Donor | | | | | | | |
|---|---|---|---|---|---|---|---|---|---|
| | | $P_1$ | $P_2$ | $P_3$ | $P_4$ | $P_5$ | $P_6$ | $P_7$ | $P_8$ |
| | | O⁻ | O⁺ | A⁻ | A⁺ | B⁻ | B⁺ | AB⁻ | AB⁺ |
| $P_1$ | O⁻ | √ | | | | | | | |
| $P_2$ | O⁺ | √ | √ | | | | | | |
| $P_3$ | A⁻ | √ | | √ | | | | | |
| $P_4$ | A⁺ | √ | √ | √ | √ | | | | |
| $P_5$ | B⁻ | √ | | | | √ | | | |
| $P_6$ | B⁺ | √ | √ | | | √ | √ | | |
| $P_7$ | AB⁻ | √ | | √ | | √ | | √ | |
| $P_8$ | AB⁺ | √ | √ | √ | √ | √ | √ | √ | √ |

## 2. Literature Review

Blood products depending upon voluntary donation are invaluable products having inherent unpredictability in supplies and demands, and lives could be lost if this matter does not properly take into account. It is, important to maintain enough inventory. According to the 2009 U.S. national blood survey, approximately fifteen million RBCs units are transfused annually [4]. Therefore, the significance of an effective BSCM in terms of service-level KPI is the impetus drove this research to equally distribute supplies concerning medical centers' demand for the first time in the BSCM context. On the contrary, service level analysis is a well-acknowledged concept in the broader background of Supply Chain Management. A supply chain network considering the sustainability design concept employed service level under uncertain conditions [5]. The model further expanded by utilizing chance constraints. Wang et al. (2020) [6] investigated the retailers' distance and price factors that influence rivals' market share and consequently measuring the service level index.

Disruption Risk (DS) refers to a variety of uncertainties, such as power failure and uncontrolled lab equipment malfunctioning in the blood inventory facilities. Disrupted logistic processes may have adverse effects on the BSC for a while. Cheraghi and Hosseini-Motlagh (2018) [7] proposed a mixed-integer robust bi-objective formulation for disaster relief in the settings of BSCM. They proposed a three-phased approach to identify candidate locations for establishing blood facilities employing the fuzzy-Vikor method as a strategic decision in the first phase. A robust risk management optimization approach was applied in the second phase. In the third phase, a two-fold formulation was proposed. An integrated programming model was developed by [8] took into account the following aspects: (a) risk of operation and disruption at the same time, (b) blood types, and their respective shelf lives. The authors proposed a hybrid two-phase approach in which the first phase proposed a combination of FAHP-GRA technique and p-robust formulation to address the risk of disruption. In the second phase, fuzzy-robust programming was developed to address operational risk.

Stochastic programming is a branch of optimization that deals with unpredictable decision-making. A particular case is two-stage stochastic programming, for instance. Zhou et al. (2019) [9] suggested a two-stage stochastic programming model for the optimal planning of petroleum supply chain logistics networks under uncertainty. Additionally, due to a wide number of criteria conditions, the sample average approximation was adopted, as well. We organized a summary of recent research papers in Table 2.

Thus far, three key terms of our research briefly explained, the unique approach of service level in BSCM, the possibility of disruption in facilities, and finally, consideration of uncertainty in terms of the mathematical modeling approach. Indeed, Hosseini-Motlagh et al. (2020) [10] proposed an excellent model that could serve as a vehicle for our contributions, too. However, there were drawbacks that we strived to improve significantly. The followings are the main differences that distinguish our research from its basis:

- The level of donors is taken into account as the source of RBCs through drawing blood by apheresis method.

- Supplies and demands are considered uncertain, whereas in the base model demand was the only uncertain parameter.
- Blood shortage is embedded into our model, whereas simply considering all demands are satisfied unrealistically.
- Service level analysis is considered as an objective function.
- Unlike the base formulation, the inventory policy is not merely First-In-First-Out (FIFO) to reduce the number of outdated units, instead, the cross-matching opportunities are reflected in the inventory policy. Therefore, inventory equality constraints were overhauled thoroughly.

Table 2. Literature review

| Reference | Method | Solution | Features | | | Stages | | | | | | |
|---|---|---|---|---|---|---|---|---|---|---|---|---|
| | | | | | | The First Stage | | | The Second Stage | | | |
| | | | P | UD | US | L | I | D/R | L | D/R | S | W | A |
| [11] | SP | Efficient hybridized method | ✓ | ✓ | ✓ | | | | | ✓ | ✓ | ✓ | ✓ |
| [12] | Scenario-based | The multi-objective solution | | | ✓ | | | | | ✓ | ✓ | | ✓ |
| [13] | Scenario-based | Branch and Cut | ✓ | ✓ | ✓ | | | | ✓ | | ✓ | | |
| [14] | Scenario-based | Goal programming method | ✓ | ✓ | | | | | ✓ | ✓ | | | |
| [15] | Internal uncertainty | Lagrangian relaxation-based algorithm | ✓ | ✓ | | | | | | ✓ | ✓ | | ✓ |
| [16] | Fuzzy-robust | Robust possibility flexible chance constrained programming model | ✓ | ✓ | | | | | ✓ | ✓ | | | ✓ |
| [8] | Fuzzy-robust | hybrid two-phase approach | ✓ | ✓ | | | | | | ✓ | | ✓ | ✓ |
| This paper | two-stage stochastic programming | ε-constraint method | | ✓ | ✓ | ✓ | ✓ | ✓ | ✓ | ✓ | ✓ | ✓ | ✓ |

P: Perishability; UD: Uncertainty in Demand; US: Uncertainty in Supply; L: Location; I: Inventory prepositioning; D/R: Distribution flow/Routing; S: Shortage; W: Wastage; A: Assignment; SP: Stochastic Programming;

## 3. Problem Statement

This paper addresses designing and planning for RBCs supply chain network that comprises donors, bloodmobiles, blood banks, blood centers, and hospitals. The RBCs supply network is responsible for blood collection, screening, storage, processes, and eventually distribution. An instance of the RBCs network is illustrated in Figure 1. Blood donations occur in the first level of the chain. RBCs units are collected in view of the two major medical preferences, blood types, and Rh factors. As shown in Table 1, these medical preferences are itemized in a compatibility matrix. For instance, a RBCs unit type A+ might be administered to a recipient with blood type A+, AB+. A recipient of RBCs type A+ can receive the following units from A+, A, O+, O- donors.

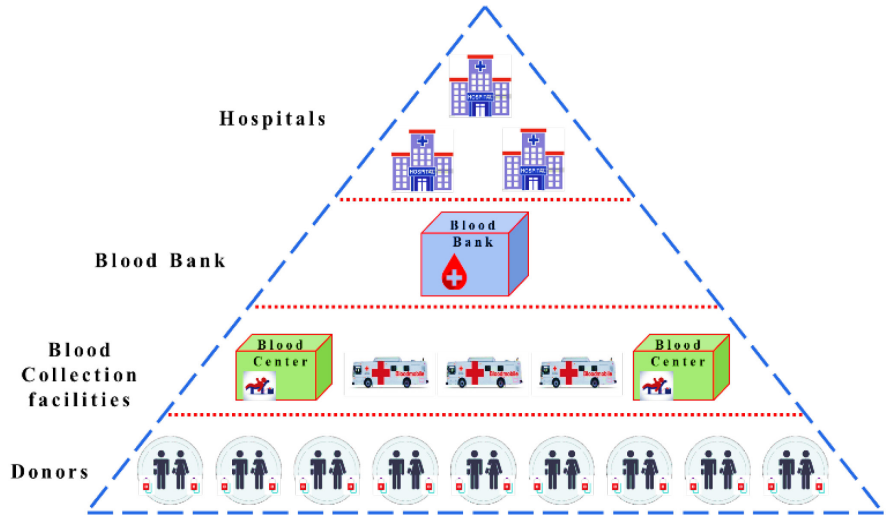

Figure 1. Blood supply chain network.

Donors contribute RBCs to either a blood center or a bloodmobile. A preliminary screening protocol commences immediately after a donor arrives to prevent infectious disease transmission. Furthermore, bloodmobiles deliver their collected RBCs units to blood centers at the end of each period. Figure 1 indicates the second stage of blood collection elements including blood centers and bloodmobiles. Blood banks are located on the third level of the network. At this level, collected blood units are carefully transported to the blood bank from blood centers where a wide range of advanced tests are examined for the units. Lastly, hospitals which are considered as demand points for RBCs units situated at the fourth level. Hospitals place orders to the assigned blood banks and are permitted to have inventories where they can administer substituted RBCs complying with medical procedures. Should hospitals have surplus units left from prior periods, crossmatching would be a worthwhile decision.

*3.1. Assumptions*

The following assumptions are made to facilitate structuring a model:
- Supplies and demands of RBCs units are considered to be uncertain.
- Another source of uncertainty is the possibility of disruption in bloodmobiles, blood centers, and blood banks.
- There is a limited number of bloodmobiles available.
- Hospitals may encounter shortages at any time or under any scenario.
- The shelf lives of RBCs units are embedded into the formulation, and the FIFO policy is considered along with crossmatching (Table 1).
- According to medical recommendation [17], it is recommended to utilize the same type of RBCs units for each patient to avoid any adverse effects of crossmatching therefore the priority rules (Table 3) are noted.

Table 3. Priority orders of RBCs [10]

| Recipient | | Priority | | | | | | | |
|---|---|---|---|---|---|---|---|---|---|
| | | 1 | 2 | 3 | 4 | 5 | 6 | 7 | 8 |
| $P_1$ | O$^-$ | $P_1$ | | | | | | | |
| $P_2$ | O$^+$ | $P_2$ | $P_1$ | | | | | | |
| $P_3$ | A$^-$ | $P_3$ | $P_1$ | | | | | | |
| $P_4$ | A$^+$ | $P_4$ | $P_3$ | $P_2$ | $P_1$ | | | | |
| $P_5$ | B$^-$ | $P_5$ | $P_1$ | | | | | | |
| $P_6$ | B$^+$ | $P_6$ | $P_5$ | $P_2$ | $P_1$ | | | | |
| $P_7$ | AB$^-$ | $P_7$ | $P_5$ | $P_3$ | $P_1$ | | | | |
| $P_8$ | AB$^+$ | $P_8$ | $P_7$ | $P_6$ | $P_5$ | $P_4$ | $P_3$ | $P_2$ | $P_1$ |

*3.2. Notations*

We provide sets of indices, parameters, and variables to formulate a two-stage stochastic programming model.

*3.2.1. Sets*
$d \in D$: Set of regions where donors are located.
$m \in M$: Set of the potential location where the bloodmobiles can be deployed.
$c \in C$: Set of existing blood centers.
$b \in B$: Set of existing blood banks.
$h \in H$: Set of hospitals which are considered as demand points.
$s^s \in S^S$: Set of supply scenarios.
$s^d \in S^d$: Set of demand scenarios.
$s^r \in S^r$: Set disruption scenarios.
$t \in T$: Set of time periods.
$\delta, \delta' \in \Delta$: Set of RBCs types.

*3.2.2. Parameters*
$O^M$: Operating fee of one RBCs unit in a bloodmobile.
$O_c^C$: Operating fee of one RBCs unit in blood center $c$.
$O_b^B$: Operating fee of one RBCs unit in blood bank $b$.
$O_h^H$: Operating fee of one RBCs unit in hospital $h$.
$H_c^C$: Holding fee of one RBCs unit in blood center $c$.
$H_b^B$: Holding fee of one RBCs unit in blood bank $b$.
$H_h^H$: Holding fee of one RBCs unit in hospital $h$.
$W_{\delta c}^C$: Wastage fee of blood type $\delta$ in blood center $c$.
$W_{\delta b}^B$: Wastage fee of blood type $\delta$ in blood bank $b$.
$W_{\delta h}^H$: Wastage fee of blood type $\delta$ in hospital $h$.
$R_{mc}^M$: Transportation fee of one RBCs unit from bloodmobile $m$ to blood center $c$.
$R_{cb}^C$: Transportation fee of one RBCs unit from blood center $c$ to blood bank $b$.
$R_{bh}^B$: Transportation fee of one RBCs unit from blood bank $b$ to hospital $h$.
$F$: Fixed cost of moving each bloodmobile between candidate locations.
$N^M$: Number of existing bloodmobiles.
$SL$: Shelf life of RBCs.
$CM_{\delta\delta'}$: Crossmatching matrix for RBCs unit type $\delta$ which is satisfied by type $\delta'$.
$PM_{\delta\delta'}$: Priority matrix for RBCs unit type $\delta$ which is satisfied by type $\delta'$.
$PF_{\delta\delta'}$: Penalty fee for RBCs unit type $\delta$ which is satisfied by type $\delta'$.
$S_{\delta dt}^{s^s}$: Supply of RBCs type $\delta$ at donor region $d$ in period $t$ under scenario $s^s$.
$D_{\delta ht}^{s^d}$: Demand of RBCs type $\delta$ at hospital $h$ in period $t$ under scenario $s^d$.
$p^{s^s}$: Probability of supply scenario $s^s$.
$p^{s^d}$: Probability of demand scenario $s^d$.
$p^{s^r}$: Probability of disruption scenario $s^r$.
$\alpha_m^{s^r}$: Percentage of Disruption in bloodmobile $m$ under scenario $s^r$.
$\alpha 1_c^{s^r}$: Percentage of Disruption in blood center $c$ under scenario $s^r$.
$\alpha 2_b^{s^r}$: Percentage of Disruption in blood bank b under scenario $s^r$.
$U_{max}$: Maximum capacity of transportation units among the network's components.
$U_c^C$: Maximum inventory capacity in blood center $c$.
$U_b^B$: Maximum inventory capacity in blood bank $b$.
$U_h^H$: Maximum inventory capacity in hospital $h$.
$\lambda$: A big number.

*3.2.3. Variables*
*First stage variable*
$z_{dct}$: Binary variable is equal to one if donor $d$ is allocated to blood center $c$ in period $t$.
$z_{dmt}^1$: Binary variable is equal to one if donor $d$ is allocated to bloodmobile $m$ in period $t$.
$z_{mct}^2$: Binary variable is equal to one if bloodmobile $m$ is allocated to blood center $c$ in period $t$.

$z^3_{cbt}$: Binary variable is equal to one if blood center $c$ is allocated to blood bank $b$ in period $t$.
*Second stage variable*
$z4^{s^s s^d s^r}_{bht}$: Binary variable is equal to one if blood bank $b$ is allocated to hospital $h$ in period $t$ under scenario $s^s$, $s^d$, and $s^r$.
$x^{s^s s^d s^r}_{\delta dmt}$: Number of RBCs units of type $\delta$ donated by donor $d$ to bloodmobile $m$ in period $t$ under scenario $s^s$, $s^d$, and $s^r$.
$x1^{s^s s^d s^r}_{\delta mct}$: Number of RBCs units of type $\delta$ transported from bloodmobile $m$ to center $c$ in period $t$ under scenario $s^s$, $s^d$, and $s^r$.
$x2^{s^s s^d s^r}_{\delta dct}$: Number of RBCs units of type $\delta$ donated by donor $d$ to blood center $c$ in period $t$ under scenario $s^s$, $s^d$, and $s^r$.
$x3^{s^s s^d s^r}_{\delta cbt}$: Number of RBCs units of type $\delta$ transported from blood center $c$ to blood bank $b$ in period $t$ under scenario $s^s$, $s^d$, and $s^r$.
$x4^{s^s s^d s^r}_{\delta bht}$: Number of RBCs units of type $\delta$ transported from blood bank $b$ to hospital $h$ in period $t$ under scenario $s^s$, $s^d$, and $s^r$.
$I^{s^s s^d s^r}_{\delta ct}$: Inventory level of RBCs units of type $\delta$ in blood center $c$ in period $t$ under scenario $s^s$, $s^d$, and $s^r$.
$I1^{s^s s^d s^r}_{\delta bt}$: Inventory level of RBCs units of type $\delta$ in blood bank $b$ in period $t$ under scenario $s^s$, $s^d$, and $s^r$.
$I2^{s^s s^d s^r}_{\delta ht}$: Inventory level of RBCs units of type $\delta$ in hospital $h$ in period $t$ under scenario $s^s$, $s^d$, and $s^r$.
$O^{s^s s^d s^r}_{\delta ct}$: Outdated units of type $\delta$ in blood center $c$ in period $t$ under scenario $s^s$, $s^d$, and $s^r$.
$O1^{s^s s^d s^r}_{\delta bt}$: Outdated units of type $\delta$ in blood bank $b$ in period $t$ under scenario $s^s$, $s^d$, and $s^r$.
$O2^{s^s s^d s^r}_{\delta ht}$: Outdated units of type $\delta$ in hospital $h$ in period $t$ under scenario $s^s$, $s^d$, and $s^r$.
$Q^{s^s s^d s^r}_{\delta ht}$: Shortages units of type $\delta$ in hospital $h$ in period $t$ under scenario $s^s$, $s^d$, and $s^r$.
$y^{s^s s^d s^r}_{\delta mt}$: Processed units of type $\delta$ in bloodmobile $m$ in period $t$ under scenario $s^s$, $s^d$, and $s^r$.
$y1^{s^s s^d s^r}_{\delta ct}$: Processed units of type $\delta$ in blood center $c$ in period $t$ under scenario $s^s$, $s^d$, and $s^r$.
$y2^{s^s s^d s^r}_{\delta bt}$: Processed units of type $\delta$ in blood bank $b$ in period $t$ under scenario $s^s$, $s^d$, and $s^r$.
$Sub^{s^s s^d s^r}_{\delta \delta' ht}$: Number of RBCs units of type $\delta$ from the inventory of hospital $h$ which is satisfied by type $\delta'$ in period $t$ under scenario $s^s$, $s^d$, and $s^r$.

### 3.3. Formulation

$$FC = F \sum_d \sum_m \sum_t z^1_{dmt} \tag{1}$$

$$OC^{s^s s^d s^r} = \sum_\delta \sum_t \left[ \left( O^M \cdot \sum_d \sum_m x^{s^s s^d s^r}_{\delta dmt} \right) + \left( \sum_d \sum_c O^C_c \cdot x2^{s^s s^d s^r}_{\delta dct} \right) + \left( \sum_m \sum_c O^C_c \cdot x1^{s^s s^d s^r}_{\delta mct} \right) \right.$$
$$\left. + \left( \sum_c \sum_b O^B_b \cdot x3^{s^s s^d s^r}_{\delta cbt} \right) + \left( \sum_b \sum_h O^H_h \cdot x4^{s^s s^d s^r}_{\delta bht} \right) \right] \tag{2}$$

$$HC^{s^s s^d s^r} = \sum_\delta \sum_t \left[ \left( \sum_c H^C_c \cdot I^{s^s s^d s^r}_{\delta ct} \right) + \left( \sum_b H^B_b \cdot I1^{s^s s^d s^r}_{\delta bt} \right) + \left( \sum_h H^H_h \cdot I2^{s^s s^d s^r}_{\delta ht} \right) \right] \tag{3}$$

$$WC^{s^s s^d s^r} = \sum_\delta \sum_t \left[ \left( \sum_c W^C_{\delta c} \cdot O^{s^s s^d s^r}_{\delta ct} \right) + \left( \sum_b W^B_{\delta b} \cdot O1^{s^s s^d s^r}_{\delta bt} \right) + \left( \sum_h W^H_{\delta h} \cdot O2^{s^s s^d s^r}_{\delta ht} \right) \right] \tag{4}$$

$$TC^{s^s s^d s^r} = \sum_\delta \sum_t \left[ \left( \sum_m \sum_c R^M_{mc} \cdot x1^{s^s s^d s^r}_{\delta mct} \right) + \left( \sum_c \sum_b R^C_{cb} \cdot x3^{s^s s^d s^r}_{\delta cbt} \right) + \left( \sum_b \sum_h R^B_{bh} \cdot x4^{s^s s^d s^r}_{\delta bht} \right) \right] \tag{5}$$

$$SubC^{s^s s^d s^r} = \sum_\delta \sum_{\delta'} PF_{\delta \delta'} \cdot PM_{\delta \delta'} \left( \sum_h \sum_t Sub^{s^s s^d s^r}_{\delta \delta' ht} \right) \tag{6}$$

$$Min\ z_1 = FC + \left[ \sum_{s^s} \sum_{s^d} \sum_{s^r} p^{s^s} \cdot p^{s^d} \cdot p^{s^r} \cdot \left( OC^{s^s s^d s^r} + HC^{s^s s^d s^r} + WC^{s^s s^d s^r} + TC^{s^s s^d s^r} + SubC^{s^s s^d s^r} \right) \right] \tag{7}$$

The first objective function, Eq. (7) minimizes the total fee of the BSC including the fixed fee of deploying bloodmobiles shown in Eq. (1). Operating fees for blood-mobiles, blood centers, blood banks, and hospitals, Eq. (2). Inventory holding fees for blood centers, blood banks, and hospitals are calculated in Eq. (3). Wastage fees for blood centers, blood banks, and hospitals are calculated in Eq. (4). Transportation fees of RBCs units are computed in Eq. (5). Finally, Eq. (6) calculates the cost of substitution.

$$Max\, z_2 = min\left\{\frac{\sum_t\left[1-\sum_{s^s}\sum_{s^d}\sum_{s^r}p^{s^s}\cdot p^{s^d}\cdot p^{s^r}\cdot \sum_\delta\left(\frac{Q^{s^s s^d s^r}_{\delta ht}}{D^{s^d}_{\delta ht}}\right)\right]}{|T|}\;\forall h\right\}$$  (8)

The second objective function, Eq. (8) intends to maximize the minimum service-level among hospitals. It adheres to the concept of service level, known as service level type β [18], in which the percentage of demand fulfilled for a hospital over the planning horizon.

$$\sum_c z_{dct} + \sum_m z^1_{dmt} \leq 1,\;\forall d,t$$  (9)

The first constraint, Eq. (9), guarantees that each donor is at most assigned either to a center or a bloodmobile in each period.

$$\sum_d\sum_m z^1_{dmt} \leq N^M,\;\forall t$$  (10)

$$\sum_d z^1_{dmt} \leq 1,\;\forall m,t$$  (11)

Eq. (10) allows the number of available bloodmobiles to deploy only at each peri-od. Eq. (11) ensure that each bloodmobile is assigned to one donor and each period.

$$\sum_\delta x^{1 s^s s^d s^r}_{\delta mct} \leq U^C_c \cdot z^2_{mct},\;\forall m,c,t,s^s,s^d,s^r$$  (12)

$$\sum_\delta x^{3 s^s s^d s^r}_{\delta cbt} \leq U^B_b \cdot z^3_{cbt},\;\forall c,b,t,s^s,s^d,s^r$$  (13)

$$\sum_c z^2_{mct} \leq 1,\;\forall m,t$$  (14)

$$\sum_b z^3_{cbt} \leq 1,\;\forall c,t$$  (15)

$$\sum_c x^{1 s^s s^d s^r}_{\delta mct} \leq \sum_d x^{s^s s^d s^r}_{\delta dmt},\;\forall \delta,m,t,s^s,s^d,s^r$$  (16)

Constraint (12) to (15) state the assignment of each bloodmobile to one blood center and blood center to one blood bank, respectively with consideration of the maximum associated inventory under the occurrence of any scenario. Moreover Eq. (16) states the amount of delivered blood by bloodmobile is less than what it collects.

$$\sum_b z^{4 s^s s^d s^r}_{bht} \leq 1,\;\forall h,t,s^s,s^d,s^r$$  (17)

$$\sum_\delta x^{4 s^s s^d s^r}_{\delta bht} \leq U^H_h \cdot z^{4 s^s s^d s^r}_{bht},\;\forall b,h,t,s^s,s^d,s^r$$  (18)

Eq. (17) and (18) ensure that each hospital is assigned to at most one blood bank with associated hospital's inventory under the occurrence of any scenario.

$$\left(1-\alpha^{s^r}_m\right)\cdot\sum_d x^{s^s s^d s^r}_{\delta dmt} = y^{s^s s^d s^r}_{\delta mt},\;\forall \delta,m,t,s^s,s^d,s^r$$  (19)

$$\sum_c x^{1 s^s s^d s^r}_{\delta mct} \leq y^{s^s s^d s^r}_{\delta mt},\;\forall \delta,m,t,s^s,s^d,s^r$$  (20)

$$\sum_m x^{1 s^s s^d s^r}_{\delta mct} + \left(1-\alpha^{1 s^r}_c\right)\cdot\sum_d x^{2 s^s s^d s^r}_{\delta dct} = y^{1 s^s s^d s^r}_{\delta ct},\;\forall \delta,c,t,s^s,s^d,s^r$$  (21)

$$\sum_b x^{3s^s s^d s^r}_{\delta cbt} \leq y^{1s^s s^d s^r}_{\delta ct}, \quad \forall \delta, c, t, s^s, s^d, s^r \tag{22}$$

$$\left(1 - \alpha^{2s^r}_b\right) \cdot \sum_c x^{3s^s s^d s^r}_{\delta cbt} = y^{2s^s s^d s^r}_{\delta bt}, \quad \forall \delta, b, t, s^s s^d s^r \tag{23}$$

$$\sum_h x^{4s^s s^d s^r}_{\delta bht} \leq y^{2s^s s^d s^r}_{\delta bt}, \quad \forall \delta, b, t, s^s s^d s^r \tag{24}$$

Constraints (19), (21), and (23) calculates the processed number of RBCs unit at bloodmobiles, blood center, and blood banks, respectively, under each scenario. Constraints (20), (22), and (24) limit the number of processed RBCs units transported from one level of chain to the other under each scenario

$$x^{1s^s s^d s^r}_{\delta dmt} \leq S^{s^s}_{\delta dt} \cdot z^1_{dmt}, \quad \forall \delta, d, m, t, s^s s^d s^r \tag{25}$$

$$x^{2s^s s^d s^r}_{\delta dct} \leq S^{s^s}_{\delta dt} \cdot z_{dct}, \quad \forall \delta, d, c, t, s^s s^d s^r \tag{26}$$

Constraints (25), and (26) assign the supplies from donors to the bloodmobiles and blood centers.

$$O^{s^s s^d s^r}_{\delta ct} = \max\left\{0, I^{s^s s^d s^r}_{\delta c,(t-SL)} - \sum_b x^{3s^s s^d s^r}_{\delta cbt} - O^{s^s s^d s^r}_{\delta c,(t-SL)}\right\}, \quad \forall \delta, c, s^s s^d s^r, t : t \geq SL+1 \tag{27}$$

$$O^{1s^s s^d s^r}_{\delta bt} = \max\left\{0, I^{1s^s s^d s^r}_{\delta b,(t-SL)} - \sum_h x^{4s^s s^d s^r}_{\delta bht} - O^{1s^s s^d s^r}_{\delta b,(t-SL)}\right\}, \quad \forall \delta, b, s^s s^d s^r, t : t \geq SL+1 \tag{28}$$

$$O^{2s^s s^d s^r}_{\delta ht} = \max\left\{0, I^{2s^s s^d s^r}_{\delta h,(t-SL)} - \sum_{\delta'} Sub^{s^s s^d s^r}_{\delta \delta' ht} - O^{2s^s s^d s^r}_{\delta h,(t-SL)}\right\}, \quad \forall \delta, h, s^s s^d s^r, t : t \geq SL+1 \tag{29}$$

Constraints (27), (28), and (29) guarantee that FIFO policy is applied in inventory management by considering their shelf lives and substituted RBCs units.

$$I^{s^s s^d s^r}_{\delta c(t-1)} + \sum_m x^{1s^s s^d s^r}_{\delta mct} + \sum_d x^{2s^s s^d s^r}_{\delta dct} - O^{s^s s^d s^r}_{\delta ct} - \sum_b x^{3s^s s^d s^r}_{\delta cbt} = I^{s^s s^d s^r}_{\delta ct}, \quad \forall \delta, c, t, s^s s^d s^r \tag{30}$$

$$I^{1s^s s^d s^r}_{\delta b(t-1)} + \sum_c x^{3s^s s^d s^r}_{\delta cbt} - O^{1s^s s^d s^r}_{\delta bt} - \sum_h x^{4s^s s^d s^r}_{\delta bht} = I^{1s^s s^d s^r}_{\delta bt}, \quad \forall \delta, b, t, s^s s^d s^r \tag{31}$$

Equality constraints (30), and (31) state the inventory constraints for blood centers and blood banks.

$$\sum_b x^{4s^s s^d s^r}_{\delta bht} + I^{2s^s s^d s^r}_{\delta h(t-1)} + \sum_{\delta'} Sub^{s^s s^d s^r}_{\delta \delta' ht} - \sum_{\delta'} Sub^{s^s s^d s^r}_{\delta' \delta ht} - O^{2s^s s^d s^r}_{\delta ht}$$
$$= D^{s^d}_{\delta ht} + I^{2s^s s^d s^r}_{\delta ht} - Q^{s^s s^d s^r}_{\delta ht}, \quad \forall \delta, h, t, s^s, s^d, s^r, \delta \neq \delta' \tag{32}$$

$$Q^{s^s s^d s^r}_{\delta ht} \cdot I^{2s^s s^d s^r}_{\delta ht} = 0, \quad \forall \delta, h, t, s^s s^d s^r \tag{33}$$

$$\left(\sum_{\delta'} Sub^{s^s s^d s^r}_{\delta \delta' ht}\right) \cdot I^{2s^s s^d s^r}_{\delta ht} = 0, \quad \forall \delta, h, t, s^s s^d s^r \tag{34}$$

Equality constraint, Eq. (32), implies that the received RBCs from blood banks or hospital's prior inventory, crossmatching units (in two forms), and outdated units are equal to the demand, inventory, and shortages. Additionally, Eq. (33) prevents inventory and shortage at the same time. Eq. (34) prevents administering substitutions if there are any availabilities of the required one. (Nonlinear equation could be linearized by techniques provided by [19].)

$$\sum_{\delta,(if\ CM_{\delta\delta'}=1)} Sub^{s^s s^d s^r}_{\delta \delta' ht} \leq I^{2s^s s^d s^r}_{\delta' h(t-1)}, \quad \forall \delta', h, t, s^s, s^d, s^r, \delta \neq \delta' \tag{35}$$

Equality (35) computes the number of RBCs units used as substitutions to satisfy a share of demand according to the prior hospital's inventory level.

$$\sum_{\delta} I^{s^s s^d s^r}_{\delta ct} \leq U^C_c, \quad \forall c,t,s^s s^d s^r \tag{36}$$

$$\sum_{\delta} I^{1 s^s s^d s^r}_{\delta bt} \leq U^B_b, \quad \forall b,t,s^s s^d s^r \tag{37}$$

$$\sum_{\delta} I^{2 s^s s^d s^r}_{\delta ht} \leq U^H_h, \quad \forall h,t,s^s s^d s^r \tag{38}$$

Constraints (36), (37), and (38) denote the inventory capacity of each blood center, blood bank, and hospital in every period and under each scenario.

$$\sum_{\delta} x^{1 s^s s^d s^r}_{\delta mct} \leq z^2_{mct} \cdot U_{max}, \quad \forall m,c,t,s^s s^d s^r \tag{39}$$

$$\sum_{\delta} x^{3 s^s s^d s^r}_{\delta cbt} \leq z^3_{cbt} \cdot U_{max}, \quad \forall c,b,t,s^s s^d s^r \tag{40}$$

$$\sum_{\delta} x^{4 s^s s^d s^r}_{\delta bht} \leq z^4_{bht} \cdot U_{max}, \quad \forall b,h,t,s^s s^d s^r \tag{41}$$

Constraints (39), (40), and (41) limit the number of transported blood units among each level.

$$z_{dct}, z^1_{dmt}, z^2_{mct}, z^3_{cbt}, z^{4 s^s s^d s^r}_{bht} \in \{0,1\}, \quad \forall d,m,c,b,h,t,s^s s^d s^r \tag{42}$$

$$\left\{ \begin{array}{l} x^{s^s s^d s^r}_{\delta dmt}, x^{1 s^s s^d s^r}_{\delta mct}, x^{2 s^s s^d s^r}_{\delta dct}, x^{3 s^s s^d s^r}_{\delta cbt}, x^{4 s^s s^d s^r}_{\delta bht}, I^{s^s s^d s^r}_{\delta ct}, I^{1 s^s s^d s^r}_{\delta bt}, I^{2 s^s s^d s^r}_{\delta ht} \\ O^{s^s s^d s^r}_{\delta ct}, O^{1 s^s s^d s^r}_{\delta bt}, O^{2 s^s s^d s^r}_{\delta ht}, y^{s^s s^d s^r}_{\delta mt}, y^{1 s^s s^d s^r}_{\delta ct}, y^{2 s^s s^d s^r}_{\delta bt}, Q^{s^s s^d s^r}_{\delta ht}, Sub^{s^s s^d s^r}_{\delta \delta' ht} \end{array} \right\} \in Z, \quad \forall \delta, \delta', d, m, c, b, h, t, s^s, s^d, s^r \tag{43}$$

Constraints (42) and (43) imply the domains of decision variables.

## 4. Toy-Example

To validate the proposed model, a toy-example is catered for the interested readers. Herein states the number of facilities in the four-level BSC. We consider six donor regions, two bloodmobiles, one blood center, one blood bank, and two hospitals. There are two scenarios with equal probability of occurrence and five time periods. Although it might seem a rather small problem size, we observe the following statistics: 19391 variables, 8937 inequality constraints, and 2880 equality constraints. The problem is run on a personal laptop with Intel Core i5 CPU® and 12 GB of RAM coded on Matlab 2019 IDE software with the IBM ILOG MILP solver package version 12.8. Since the concentration of this paper is the study of service level, we only pre-sent the detailed information of the hospital no. one. Figure 2 depicts the demand of hospital no. one in terms of the total number of RBCs units. Therefore, number 300, which is demonstrated in the first column and first row, is the total number of RBCs units required in the first period under the occurrence of the first scenario.

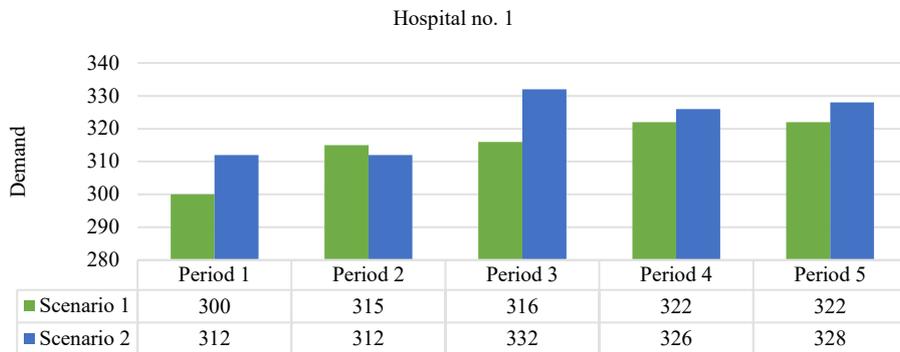

Figure 2. RBCs demand units for hospital no. 1

Figure 3, however, depicts the total transported number of RBCs units from the only blood bank to the hospital no. one. Therefore, number 280, which is demonstrated in the first column and first row, is the total number of RBCs units transferred in the first period under the occurrence of the first scenario.

As mentioned earlier, one of our interests lay on the reality of the problem statement by considering shortages. Therefore, there are possibilities that demand and supply are not balanced. Figure 4 numerically illustrates the subtraction of demands from supplies under the two equally probable scenarios. If the subtraction represents a positive number, a shortage happened (the total required number of RBCs units is greater than the supplied ones). Otherwise, the hospital encounters surplus. For in-stance, consider the third period while the first scenario happened, ten RBCs units are extra and preserved as inventory. However, in the same period, while the second scenario happened, the hospital fell short of eleven RBCs units.

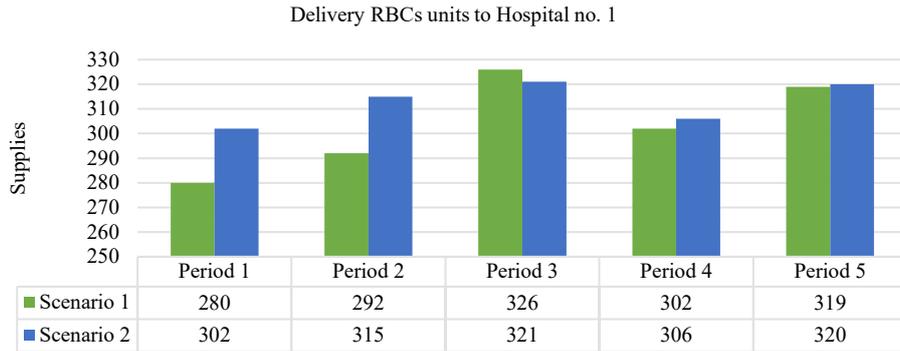

Figure 3. Delivered RBCs units from blood bank to hospital no. 1

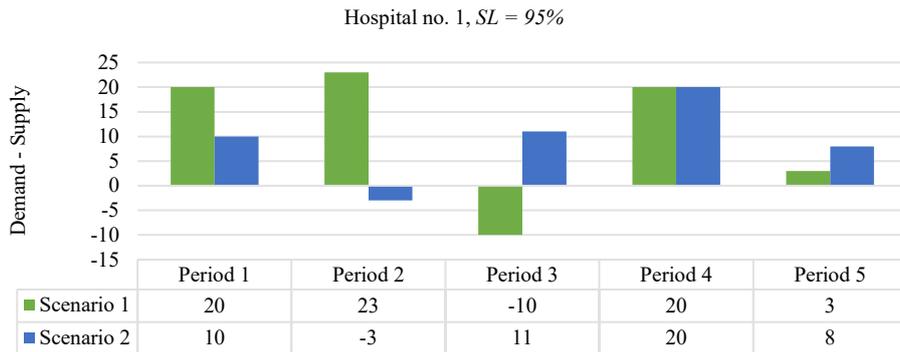

Figure 4. Difference between demand and supply for hospital no.1

Naturally, the number of required RBCs units are more than supplies, and our ex-ample also accepted that notion. Despite that fact, the computed service level for the hospital no. one is 95 %. Hence 95 % of the demand was met through our proposed model. Referring to Eq. (32), the equality constraint displays the balance between received units and demand units. On the left side of the equation, the following are considered; received RBCs units by the dedicated blood bank, previous inventory level, number of substituted RBCs units, level of inventory utilized for substitution of other RBCs type, and finally outdated units. On the right side of the equation, the following are considered; demand, excess of RBCs unit in the form of inventory at the end of the period, and shortage units. Let us consider the first scenario that happened in periods one, two, and three. Additionally, we presumed the second scenario happened for periods four and five. The hospital encountered a shortage for the first two periods and in period three, ten extra blood units are available. One instance of the Eq. (32) could be as presented in Eq. (44).

In Eq. (44), 306 RBCs units are provided, 10 units are carried over by the inventory level in the third period, 2 units received as substitution, 4 units are given as substitution, and zero units are outdated, subsequently, 314 units are the availabilities. On the right side of Eq. (44), demand is 326 units, 12 units are the shortages, and there-fore, the hospital could not preserve any units in the form of inventory.

$$\underbrace{\sum_b x^{4 s^s s^d s^r}_{\delta bht}}_{306} + \underbrace{I^{2 s^s s^d s^r}_{\delta h(t-1)}}_{10} + \underbrace{\sum_{\delta'} Sub^{s^s s^d s^r}_{\delta \delta' ht}}_{2} - \underbrace{\sum_{\delta'} Sub^{s^s s^d s^r}_{\delta' \delta ht}}_{4} - \underbrace{O^{2 s^s s^d s^r}_{\delta ht}}_{0}$$
$$= \underbrace{D^{s^d}_{\delta ht}}_{326} + \underbrace{I^{2 s^s s^d s^r}_{\delta ht}}_{0} - \underbrace{Q^{s^s s^d s^r}_{\delta ht}}_{12}, \text{ for period } (4) \tag{44}$$

## 5. Conclusion and Insights

In this paper, we presented a two-stage stochastic programming model for the BSCM by considering two central concentrations of facility disruptions and service level. Uncertainties in supplies and demands are taken into account. The model incorporated essential decisions such as blood distribution, inventory levels, crossmatching units, shortage, and finally outdated units. The model is designed to minimize the total cost of BSC while the conflicting objective of service level maximization was measured. The two-stage stochastic programming was exploited to tackle the uncertainties. Several research paths for this paper could be highlighted. The L-Shaped method is a renowned solution methodology practiced by [20] and [21] could be employed for medium- and large-scale problems. Extensive use of fuzzy numbers is recommended as an alternative to stochastic parameters, the instances are [22–24].